\documentclass{jfm}
\usepackage{graphicx}
\usepackage{epstopdf, epsfig}
\usepackage{upgreek}
\usepackage{mathtools}
\usepackage{amsmath} 

\usepackage{color}

 \newcommand{\be}{\begin{equation}}
 \newcommand{\ee}{\end{equation}}
 \newcommand{\bea}{\begin{eqnarray}}
 \newcommand{\eea}{\end{eqnarray}}

\shorttitle{Droplet jumping by modulated electrowetting}
\shortauthor{Q. Vo and T. Tran}

\title{Droplet jumping by modulated electrowetting}

\author{Quoc Vo\aff{1}
\corresp{\email{xuv1@pitt.edu}},
\and Tuan Tran\aff{2} 
\corresp{\email{ttran@ntu.edu.sg}}}

\affiliation{\aff{1}School of Medicine, 
	University of Pittsburgh, 
	4200 Fifth Ave, Pittsburgh, PA 15260, USA.
\aff{2}School of Mechanical \& Aerospace Engineering, 
	Nanyang Technological University, 
	50 Nanyang Avenue, 639798, Singapore.}
	
\begin{document}

\maketitle

\begin{abstract}
We investigate 
jumping of sessile droplets
from a solid surface in 
ambient oil using modulated electrowetting actuation.
We focus on the case in which the electrowetting effect is activated 
to cause droplet spreading and then deactivated 
exactly at the moment the droplet reaches its maximum deformation. 
By systematically varying the control parameters such as
the droplet radius, liquid viscosity, and applied voltage, 
we provide detailed characterisation 
of the resulting behaviours 
including 
a comprehensive phase diagram separating
detachment from non-detachment behaviours, as well as
how the detach velocity 
and detach time, i.e, duration leading to detachment, 
depend on the control parameters. 
We then 
construct a theoretical model
predicting the detachment condition 
using energy conservation principles.
We finally validate our theoretical analysis
by experimental data obtained in the explored ranges of
the control parameters.
\end{abstract}

\begin{keywords}
Authors should not enter keywords on the manuscript, as these must be chosen by the author during the online submission process and will then be added during the typesetting process (see http://journals.cambridge.org/data/\linebreak[3]relatedlink/jfm-\linebreak[3]keywords.pdf for the full list)
\end{keywords}

\section{Introduction}
A sessile droplet
on a flat dielectric-coated electrode 
wets the substrate more 
when a voltage is applied 
between the droplet and the electrode.
The so-called
electrowetting-on-dielectric (EWOD) phenomenon
has been rapidly 
explored in recent years 
in both fundamental research
and industrial applications.
Spreading of droplets
can be well controlled using
the EWOD effect, making
it an ideal tool to study
contact-line dynamics 
including 
wetting, dewetting \citep{Hong2014,Vo2018a},
contact angle hysteresis \citep{Nelson2011c,Sawane2015},
droplet and soft-surfaces interactions \citep{Dey2019a},
as well as
coalescence-induced jumping droplets \citep{Vahabi2018,Boreyko2009,Farokhirad2015}.
Moreover,
EWOD has been emerging as 
a powerful technique in 
various industrial applications 
such as droplet manipulation
\citep{Fair2001,Fair2007,Pollack2000a},
optical imaging systems \citep{Berge2000,Hao2014,Kuiper2004,Lee2019c}, 
liquid deposition \citep{Baret2006,Leichl2007},
and energy harvesting systems \citep{Moon2013,Xu2020}.
Understanding of droplet-substrate interactions
under electrowetting effect also helps 
design and optimisation of advanced surfaces 
such as anti-icing \citep{Mishchenko2010} and 
self-cleaning surfaces \citep{Blossey2003}.

Among applications utilising 
the electrowetting effect,
induction of droplet detachment 
from a solid surface \citep{Vo2019,Lee2014,Wang2020,Weng2021,Xiao2021} 
is not only a topic of fundamental interests,
but also a versatile tool in 
enabling manipulation of droplets in three-dimensional settings \citep{He2021,Hong2015a}.
The principle of this technique 
relies on conversion of
electrical energy to 
droplet surface energy 
by overstretching the droplet
when the electrowetting 
effect is activated. 
Subsequently, when 
the electrowetting effect 
is deactivated,
the excess surface energy of 
the overstretched droplet
converts to 
kinetic energy 
inducing droplet retraction and
detachment \citep{Vo2019}.
The critical condition 
of droplet detachment 
using electrowetting actuation
is that the excess surface energy 
of the overstretched droplet
overcomes the sum of viscous dissipation 
and elastic energy of 
the contact line during droplet retraction \citep{Vo2019}.

In practice, 
beside the applied voltage used
to control the strength 
of the electrowetting effect, 
the activating duration of 
the voltage $T_{\rm p}$
(see Fig.~\ref{fig:principle}a, top panel) 
is also an important parameter 
determining 
jumping behaviours of the 
actuated droplets.
Two major approaches 
to modulate $T_{\rm p}$
to induce droplet jumping 
by electrowetting are
actuation from equilibriated state (AES) 
and actuation from maximum 
deformation state (AMS).
In the AES approach 
(Fig.~\ref{fig:principle}a, middle panel),
the activating duration 
$T_{\rm p}$ is held
until the droplet 
reaches its new equilibrium.
In other words, $T_{\rm p}$ 
is set $\ge \uptau_{\rm e}$,
where $\uptau_{\rm e}$ is the time 
required for droplet to reach 
the new equilibrium  
after the electrowetting effect is activated \citep{Cavalli2016,Vo2019,Wang2020}.
As $\uptau_{\rm e}$
is well-defined
using system's parameters \citep{Vo2018},
AES approach is simple 
and well-controlled.
However, the ability 
to induce droplet jumping 
using the AES method 
is limited by contact angle saturation, 
i.e., the electrowetting effect
saturates 
at high applied voltage 
\citep{Mugele2005} 
resulting in a saturation 
of the excess surface energy 
supporting droplet detachment.
In contrast, 
in the AMS approach (Fig.~\ref{fig:principle}a, bottom panel),
$T_{\rm p}$ is set equal to $\uptau_{\rm m}$, i.e., 
the time for droplet 
to reach maximum deformation state  
after the electrowetting effect is activated
\citep{JunLee2012, Lee2014}.
As the excess surface energy
at the maximum deformation state 
is higher than that 
at the equilibriated state,
AMS is a more effective 
approach compared to AES.
For instance, 
for the same actuating system, 
the critical voltage 
for 
droplet detachment to occur
from AMS is lower than 
that from AES \citep{Lee2014,Wang2017}.
As a result, AMS
is a more favourable method 
in practical applications \citep{Hong2015a,Hong2015b}.
Nevertheless, 
there is yet a parametric study 
on the dynamical behaviours of 
droplet detachment using AMS.
Moreover, the critical conditions for
droplet detachment using AMS
is not yet established. 
This 
tremendously limits 
the applicability of 
the AMS method.

In this paper, 
we systematically investigate  
detachment behaviours of droplets 
under electrowetting actuation using 
AMS method by
varying three control parameters: applied voltage $U$, 
droplet's viscosity $\mu$, 
and droplet radius $r_0$.
We first confirm that AMS 
is a more effective approach
to induce droplet detachment 
compared to AES
shown by both lower 
critical detachment voltage 
and higher maximum jumping height.
We then construct a 
comprehensive phase diagram
separating detachable and nondetachable
behaviours of droplets under AMS 
varying all the three control parameters.
The dependence of detach velocity 
and detach time
on the control parameters 
are also examined in details.
Finally, we theoretically develop 
and experimentally verify a model 
describing the critical condition 
for droplet detachment using AMS approach.

\section{Experimental method}
\subsection{Experimental setup and Materials}
\begin{figure}
\includegraphics[width=0.9\textwidth]{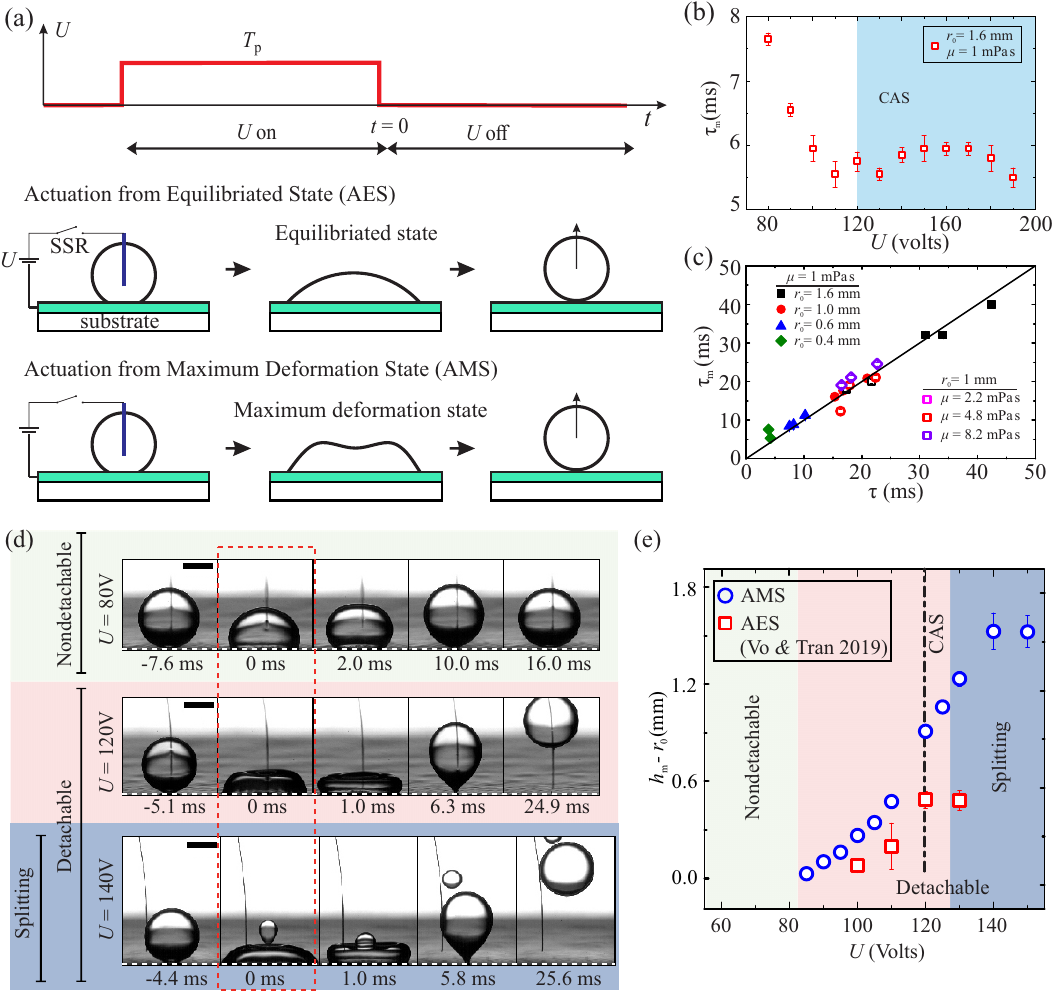}
\centering
\caption{
(a) Top panel: schematic illustrating
modulation of the applied voltage. 
The duration of the voltage pulse is $T_p$
and the time reference ($t = 0$) is set at the end of the pulse.  
Middle panel: schematic illustrating  
actuation 
from equilibrated state (AES). 
The voltage is turned on and maintained until the droplet reaches its
equilibrated state; 
subsequently the voltage is turned off, causing droplet retraction.  
Bottom panel: schematic 
illustrating actuation from maximum deformation state (AMS).
The voltage is turned on, causing a droplet to spread to its maximum deformation
state, and is then turned off exactly 
at this moment. 
(b) Plot showing the 
time to reach maximum deformation, 
$\uptau_{\rm m}$ versus $U$.
(c) Plot showing 
$\uptau_{\rm m}$ versus 
the underdamped characteristic 
spreading time
$\uptau = \pi (\rho r_0^3/\eta \sigma)^{1/2}$. 
The solid line represents 
the relation $\uptau_{\rm m} = \uptau$.
(d) Snapshots showing 
the behavioural change of a water droplet with radius $0.5\,$mm
using AMS 
when $U$ is increased.
The scalebars represent $0.5\,$mm.
(e) Plot showing the dependence of 
$h_{\rm m} - r_0$
on $U$ for 
droplets jumping from
AMS (blue circles) 
and AES (red squares).
Here, the droplet has radius $r_0 = 0.5\,$mm 
and viscosity $\mu = 1.0\,$mPa\,s; $U_{\rm s} = 120\,$V;
the experiment was done 
in 2\,cSt silicone oil.}
\label{fig:principle}
\end{figure}

To induce the electrowetting effect,
we use a substrate
made from 
an indium-tin-oxide (ITO) 
glass slide spin-coated with a layer of 
fluoropolymer (Teflon-1601, DuPont)
\citep{Vo2021c}.
We set the thickness 
of the Teflon layer 
at $d = 2.5\,{\rm \upmu m}$ 
to ensure electrical insulation
for the ITO electrode.
To apply a voltage between 
a droplet deposited on the substrate 
and the ITO electrode,
we use an $18\,\upmu$m
diameter tungsten wire
dipped into the droplet bulk
and connect it 
to the positive 
terminal of a direct-current (DC) power supply via
a normally open solid state relay (SSR) 
(see Fig.~\ref{fig:principle}a).
The negative terminal of 
the power supply is connected to
the ITO electrode.
We use a 
function generator to close the SSR, thereby
applying a voltage $U$ 
in the range $60\,{\rm V} \le U \le 200\,{\rm V}$ 
between the electrodes
for a controlled duration $T_{\rm p}$ (Fig.~\ref{fig:principle}a)
to induce the electrowetting effect and droplet jumping.
For our electrowetting substrates, 
the voltage 
causing contact angle saturation is 
$U_{\rm s} \approx 110\,\pm 10$V (Tab.~\ref{tb:1}).

\begin{table}
  \begin{center}
\def~{\hphantom{0}}
  \begin{tabular}{l c c c c}
     \%wt glycerol & $\mu$\,(mPa$\cdot$s) & $\rho$\,(${\rm kg.m}^{-3}$) & $\sigma$\,(${\rm mN.m}^{-1}$) & $U_{\rm s}$\,(V)\\[3pt]
       0 & 1.0 & 1000  & $37.2 \pm 0.5$ & $\approx 120$\\
       20 & 2.2 & 1053  &	$32.4 \pm 1.0$ & $\approx 115$\\
       41.5 & 4.8 & 1106 & $30.4 \pm 0.9$ & $\approx 110$\\
       55 & 8.2 & 1141  & $29.7 \pm 0.5$ & $\approx 110$\\
       67 & 17.6 & 1174  & $26.9 \pm 0.4$ & $\approx 105$\\
       74 & 32.7 & 1192 & $25.9 \pm 0.5$ & $\approx 100$\\
       80 & 68.7 & 1210 & $26.8 \pm 0.4$ & $\approx 105$\\
  \end{tabular}
  \caption{Measured values of viscosity $\mu$, density $\rho$ of glycerin 
solutions, and interfacial tensions $\sigma$ 
between glycerin solutions and 2\,cSt silicone oil. The CAS voltage $U_{\rm s}$ is determined experimentally by examining 
the saturation of the equilibrated contact angle 
$\theta_{\rm e}$ when varying $U$ \citep{Vo2019}.}
  \label{tb:1}
  \end{center}
\end{table}

We use aqueous glycerin 
solutions consisting of
glycerol, DI water, and 
0.125\,M sodium chloride 
to generate droplets.
The electrical conductivity of the working liquid 
is measure experimentally at $\approx 8.8 \times 10^{-4}\,{\rm S\,m^{-1}}$.
In our analysis and similar to other studies of electrowetting \citep{Mugele2005,Baret2006},
we consider liquid 
droplets perfectly conductive
and neglect the minute effect of varying liquid permittivity.
The viscosity $\mu$ of glycerin solutions
is varied
from $1.0\,{\rm mPa s}$ 
to $68.7\,{\rm mPa s}$ 
by adjusting the glycerol concentration (Tab.~\ref{tb:1}).
The droplet radius $r_0$ 
is also varied between 
$80\,{\rm \mu m}$ and $1.5\,{\rm mm}$.
We immerse 
the substrate in a pool of 
silicone oil; 
the oil's temperature is kept at $20 \pm 0.5\,^{\circ}{\rm C}$ 
to maintain consistent experimental conditions.
The use of silicone oil as the outer phase 
in our experiment is not only 
to reduce contact angle hysteresis 
but also to increase initial contact angle 
of liquid droplets on the substrates \citep{Baret2006,Hong2015b}.
The contact angle of glycerin solution droplets
deposited  
on the substrate
in the silicone oils is 
$\theta_{\rm 0} = 160^{\circ}$ 
in the absence of the electrowetting effect.
For simplicity,
the viscosity of the oil is kept fixed at 
$\mu_{\rm o} = 1.8\,{\rm mPa s}$.
Other properties of the working liquids, 
including density $\rho$ and
interfacial tension $\sigma$,
are 
measured experimentally and
given in Tab.~\ref{tb:1}.

To capture the behaviours of 
droplets under electrowetting actuation,
we use a high speed camera (Photron, SAX2), 
typically running at 5000 frames per second (FPS).
The recorded images are processed using MATLAB 
to extract the
contact radius $r$ and 
dynamic contact angle $\theta_{\rm t}$ during actuation, 
as well as the jumping height $h$ of droplets after actuation.
The measurements of $r$ and $\theta_{\rm t}$, 
as well as the uncertainty analysis, follow the same 
experimental procedure described in our previous study \citep{Vo2019}.
The uncertainty of contact angle measurements is estimated 
within $2.5^\circ$.
For each set of the control parameters ($r_0$, $\mu$, $U$),
the experiment is repeated three times.

\subsection{Electrowetting actuation}
To induce jumping of droplets using electrowetting actuation,
we note that the 
activating duration $T_{\rm p}$
plays a crucial role. 
Experimental studies have pointed to 
the time to reach maximum deformation $\uptau_{\rm m}$
as the most optimal time duration for 
jumping droplets, i.e., causing 
highest jumping \citep{Wang2017}. 
We also note that 
the dynamics of droplets actuated by 
the electrowetting effect is either underdamped
or overdamped, 
i.e., the electrowetting-induced driving force
is opposed dominantly by either 
the droplet inertia or contact-line friction, respectively \citep{Vo2018}. 
Each one of these behaviours are characterised 
by a distinct spreading 
timescales.
As it was shown that 
electrowetting-induced
droplet detachment from solid substrates
is only possible 
for spreading in the 
underdamped regime,
we set the activating duration $T_{\rm p}$
equal to the underdamped characteristic spreading time 
$\uptau = \pi(\rho r_0^3/\eta \sigma)^{1/2}$ \citep{Vo2018,Wang2020,Xiao2021}:
\begin{equation}
\label{eq:Tp}
T_{\rm p} = \pi \left( \frac{\rho r_0^3}{\eta \sigma}\right)^{1/2},
\end{equation}
where the electrowetting number $\eta$, 
also known as the electrical capillary number \citep{Hassan2023,Fallah2022}, 
is defined as $\eta = \epsilon \epsilon_0 U^2/(2 d \sigma)$ for
$U \le U_{\rm s}$ and $\eta = \epsilon \epsilon_0 U_{\rm s}^2/(2 d \sigma)$ for $U > U_{\rm s}$.
Here, $\epsilon$ and $d$ 
respectively are the dielectric constant and 
thickness of the Teflon coating,
$\epsilon_0$ the permittivity of free space,
$\rho$ the droplet density,
$\sigma$ 
the droplet-oil interfacial tension, and
$U_{\rm s} \approx 110 \pm 10\,$V
the threshold voltage
above which 
contact angle saturation (CAS) occurs. 
The threshold voltage $U_{\rm s}$
is determined experimentally 
by examining 
the dependence of 
equilibrated contact angle $\theta_{\rm e}$ 
on the applied voltage $U$ \citep{Vo2021b}.
Slight fluctuation of $U_{\rm s}$ is 
within $\pm 10\,$V and occurs 
when varying the liquid's viscosity (see Tab.~\ref{tb:1}).
We note that Eq.~\ref{eq:Tp} 
implies that
$T_{\rm p}$ 
also saturates 
when $U > U_{\rm s}$
due to the 
contact angle saturation effect, consistent with 
the experimental data 
shown in Fig.~\ref{fig:principle}b.
In Fig.~\ref{fig:principle}c,
we show an excellent agreement between 
the measured values of 
the time to reach maximum deformation
and the characteristic 
spreading time $\uptau$
of droplets actuation in the underdamped regime. 
This strongly suggests that $\uptau$
can be used to describe the time to reach maximum deformation of droplets
under electrowetting actuation. 
As a result,
the activating duration $T_{\rm p}$
in our experiment is determined by Eq.~\ref{eq:Tp}, i.e., 
dependent only 
on the experimental parameters and
free of uncertainty from 
the experimental values of $\uptau_{\rm m}$.

\section{Results and Discussions}
\subsection{Droplet jumping by EWOD actuation from maximum deformation state}

In Fig.~\ref{fig:principle}d, 
we show several series of snapshots of
actuated droplets to illustrate their 
spreading and jumping dynamics.
From the top panel to the bottom one, 
the applied voltage $U$
is increased from $80\,$V to $140\,$V, 
while the droplet radius 
and viscosity are fixed at
$r_0 = 0.5\,$mm, $\mu = 1.0\,$mPa\,s, respectively. 
Generally, we observe that 
as soon as the electrowetting
effect is activated on a droplet, 
it causes droplet spreading with
an initial contact-line velocity $v_{\rm I}$
and generates on the droplet's surface
capillary waves
propagating from 
the contact line toward the apex.
At the end of the activating 
duration ($t = 0$),
the droplet immediately recoils 
and subsequently 
jumps off from the substrate 
if the applied voltage 
is sufficiently high.
For instance, in the experiment shown 
in Fig.~\ref{fig:principle}d,
droplet detachment from 
the solid substrate occurs
for $U \ge 90\,$V. 
We also observe that at
high applied voltages, 
e.g., $U = 140\,$V 
(Fig.~\ref{fig:principle}d, last panel), 
a small satellite droplet 
is ejected from the actuated droplet.
This is due to the effect of 
the strong capillary waves 
on the droplet's surface generated 
by the electrowetting effect \citep{Vo2021b}.

In the cases that an actuated droplet 
detaches from the substrate,
the detach time $T_{\rm d}$,   
measured from the 
time reference $t = 0$ 
to the moment the droplet detaches,
reduces with the applied voltage.
For instance, $T_{\rm d}$ drops from 
$6.3\,$ms to $5.8\,$ms when 
$U$ increases from 120\,V to 140\,V 
(Fig.~\ref{fig:principle}d).
Moreover, the maximum jumping height 
$h_{\rm m}$
defined as the maximum height 
of the droplet's center of mass
increases with $U$. 
For instance, 
$h_{\rm m} - r_0$ significantly
increases from $0.38\,$mm 
to $1.55\,$mm
when $U$ increases 
from 100\,V to 140\,V
(Fig.~\ref{fig:principle}d).
Comparing the maximum jumping height 
obtained in our experiment
with the one 
obtained in the case 
jumping is induced by AES (Fig.~\ref{fig:principle}e),
we observe that for the same voltage, 
$h_{\rm m}$ obtained 
from AMS (blue circles)
is consistently higher than that 
from AES (red squares).
Moreover, 
the maximum jumping height obtained from
AMS is not limited by contact angle saturation,
in contrast to that from AES.

\begin{figure}
\includegraphics[width=0.95\textwidth]{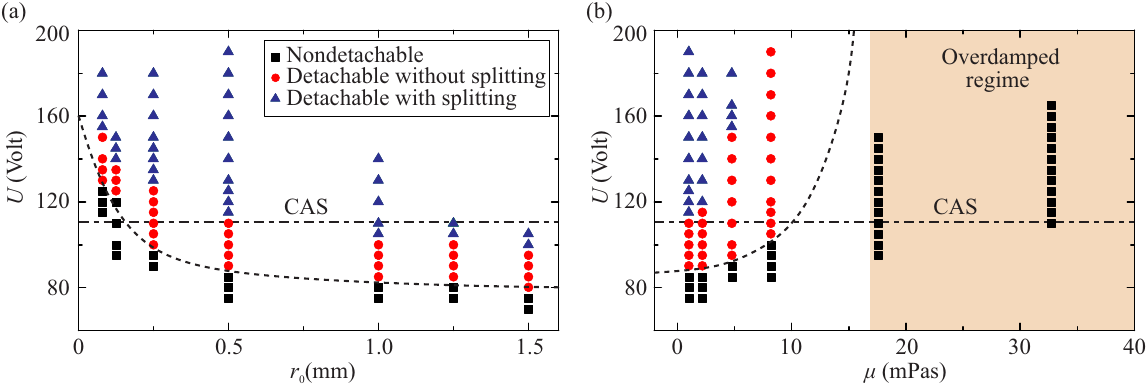}
\centering
\caption{(a) Phase diagram showing 
different behaviours of 
$\mu = 1.0$\,mPa\,s 
droplets under AMS with
varying $U$ and $r_0$. 
(b) Phase diagram showing different 
behaviours of $r_0 = 0.5$\,mm 
droplets under AMS with
varying $U$ and $\mu$.
In both diagrams, droplet behaviours are categorised into
three major regimes:
nondetachable (black squares), 
detachable without splitting (red circles)
and detachable with splitting
(blue triangles).
The dash lines are used to guide the eyes along
the boundary between the
detachable 
and the nondetachable regimes. 
The dashed-dotted lines indicates
the average contact angle saturation (CAS)
threshold 
(see Tab.~\ref{tb:1} for specific values of $U_{\rm s}$ 
at different $\mu$).}
\label{fig:phaseDiagram}
\end{figure}

In Fig.~\ref{fig:phaseDiagram},
we show the phase diagrams 
of droplet behaviours obtained 
by varying three the control parameters 
$r_0$, $\mu$, and $U$.
These observed behaviours, which 
are illustrated in Fig.~\ref{fig:principle}d,  
include nondetachable, 
detachable without splitting, 
and detachable with splitting.
The critical voltage
at the transition for detachment 
is higher 
for smaller droplet size (Fig.~\ref{fig:phaseDiagram}a)
or higher viscosity (Fig.~\ref{fig:phaseDiagram}b).
We also observe that 
the detachment of droplet 
is limited for $\mu < 17.6\,{\rm mPa\,s}$
as the transition from 
the underdamped regime 
to overdamped regime occurs at $17.6\,{\rm mPa\,s}$ 
(Fig.~\ref{fig:phaseDiagram}b) \citep{Vo2018}.
We highlight that
AMS method works for 
$r_{\rm 0}$ as small as $80\,{\rm \mu m}$, 
a limit that was not possible using
AES for the same substrate \citep{Vo2019}.

\subsection{Detach velocity and detach time}
In the case that a droplet
detaches from the solid substrate, 
the detach velocity $u_{\rm d}$, defined as 
the droplet's center-of-mass 
velocity at the moment it completely separates
from the substrate,
and the detach time $T_{\rm d}$, i.e., 
the duration from the reference time 
$t= 0$ to the separating moment, 
are the most critical parameters
representing the detaching dynamics.
Understanding of these characteristic parameters 
help optimise design and operations of 
droplet-actuating systems 
using electrowetting effect. 
In this section, we discuss 
the dependences of $u_{\rm d}$ 
and $T_{\rm d}$ on the applied voltage $U$, 
viscosity $\mu$, and droplet radius $r_{\rm 0}$.

\begin{figure}
\includegraphics[width=0.9\textwidth]{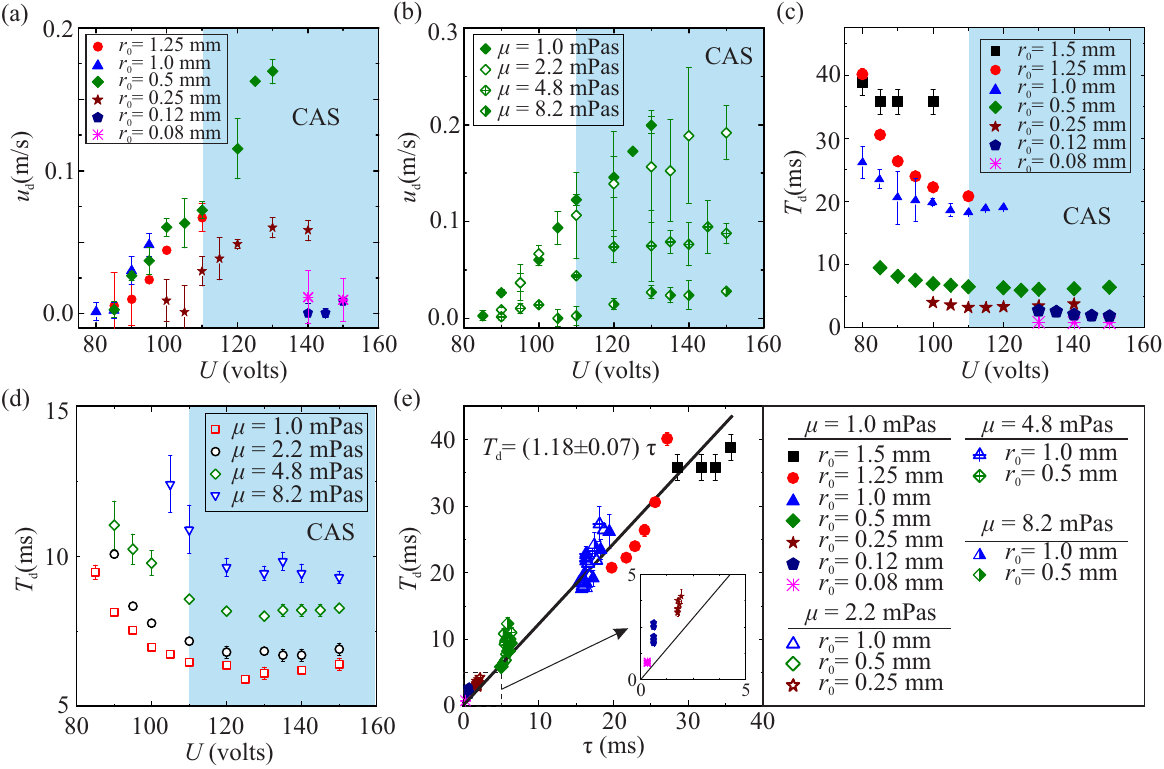}
\centering
\caption{
(a, b) Plots showing 
detach velocity $u_{\rm d}$ versus
$U$ for (a) different droplet radii $r_0$  
and (b) different droplet viscosity $\mu$. 
(c, d) Plots showing 
detach time $T_{\rm d}$ versus
$U$ for (c) different droplet radii $r_0$  
and (d) different droplet viscosity $\mu$.
(e) Plot showing 
detach time 
$T_{\rm d}$ versus $\uptau$ for various values 
of $r_0$, $U$, and $\mu$. 
Inset: a zoomed-in 
plot showing data in the dashed box.
The solid line indicates the best 
fit to the experimental 
data using the linear relation 
$T_{\rm d} = k\uptau$, 
where $k = 1.18 \pm 0.07$.
The shaded areas indicates
the average contact angle saturation (CAS)
threshold (see Tab.~\ref{tb:1} for specific values of $U_{\rm s}$ for different $\mu$).}
\label{fig:characteristic}
\end{figure}

We first focus on the 
detach velocity 
$u_{\rm d}$.
In Fig.~\ref{fig:characteristic}a,
we show the dependence of $u_{\rm d}$ 
on the applied voltage $U$ for
$r_{\rm 0}$ varying from 0.08\,mm
to 1.25\,mm and 
$\mu$ fixed at $1\,{\rm mPa\,s}$.
Generally, 
we observe that
$u_{\rm d}$
increases with $U$ as a result of 
higher electrical energy 
applied to the system.
However, 
the dependence of 
$u_{\rm d}$ on $U$ 
becomes more irregular 
at high applied voltage,
i.e., $U > 130\,$V.
For example, for $r_0 = 0.08\,$mm,
$u_{\rm d}$ varies little between $U = 140\,$V
and $U = 150\,$V.
We attribute such irregular 
dependence of
$u_{\rm d}$ on $U$ 
to hydrodynamical and electrical instabilities of
the system at high voltage.
When the applied voltage increases, 
the electrical force 
applied to the contact line 
becomes larger,
causing more abrupt and forceful deformation 
to the droplet, e.g.,
stronger capillary waves and 
even splitting of the droplet 
(Fig.~\ref{fig:principle}d).
Such violent behaviours induce 
nonlinear effects that reduce
the energy transfer efficiency,
from electrical to 
kinetic energy of the detached droplet.
Moreover, electrical leakage may 
be possible through the 
dielectric layer at high applied voltage
without breaking it down during the experiment
\citep{Moon2002}. 
This also reduces the efficiency of the 
EWOD effect in generating higher $u_{\rm d}$ 
for jumping droplets. 
Therefore, reducing irregularities at high voltage may require increasing viscosities of the liquids or using insulators with higher dielectric strength.
Next, in Fig.~\ref{fig:characteristic}b,
we plot the dependence of 
$u_{\rm d}$ on $U$ 
for $\mu$ varying
from $1\,{\rm mPa\,s}$
to $8.2\,{\rm mPa\,s}$ and $r_{\rm 0}$ fixed at $0.5\,$mm.
We observe that for $U<130\,$V, 
$u_{\rm d}$ 
linearly increases with $U$,
whereas for $U\ge130\,$V,
$u_{\rm d}$ approaches a plateau.
The increasing rate of $u_{\rm d}$
with $U$ also reduces with higher $\mu$ due to 
larger viscous dissipation.

We now examine the detach time $T_{\rm d}$.
In Fig.~\ref{fig:characteristic}c and d, 
we respectively show 
$T_{\rm d}$
vs. 
$U$ for various droplet radii 
and 
droplet viscosities.
We note that for $r_0 = 1.25\,$mm and $r_0 = 1.5\,$mm, 
the voltage is limited at 110\,V and 100\,V, respectively,
due to frequent electrical breakdowns of the 
dielectric layer separating the electrode and the liquids.
Here, electrical breakdowns increase with the actuation time, 
as well as the
contact area between the droplet
and the substrate during actuation.
Both factors increase
with larger droplet radius.

Typically, $T_{\rm d}$ decreases with 
increasing $U$ as long as the voltage is within the
contact angle saturation (CAS) limit, i.e., 
$U_{\rm s} = $110\,V in our case.
The detach time $T_{\rm d}$ eventually reaches
a plateau when the applied voltage exceeds
$U_{\rm s}$.
Here, we note that $T_{\rm d}$ for 
AMS is defined the same way 
as the droplet's retracting time,
i.e., from the moment the electrowetting effect is turned off
at the droplet's maximum deformation 
to the moment the droplet detaches from 
the surface. 
In addition, 
the retracting time of the droplet only depends on 
its maximum spreading diameter,
which is limited by CAS.
As a result, we infer that $T_{\rm d}$ also
saturates when $U \ge U_{\rm s}$,
consistent with our experimental results shown in 
Fig.~\ref{fig:characteristic}c and d.

The detach time $T_{\rm d}$, which is measured 
from the moment the voltage is released to the detaching moment, 
is defined the same way as the retracting time of the droplets from maximum deformation. 
As a result, we hypothesise that $T_{\rm d}$ is closely related to the timescale 
characterising the spreading (or retracting) dynamics of droplets. 
As the spreading droplets in our study are underdamped, 
we examine the relation between $T_{\rm d}$ 
and the underdamped characteristic spreading timescale $\uptau$ 
and show a plot of $T_{\rm d}$ vs $\uptau$ in Fig.~\ref{fig:characteristic}e. 
The plot consists of data obtained by varying 
$r_0$, $\mu$, and $U$ in their explored ranges. 
We observe that the dependence of $T_{\rm d}$ on 
$\uptau$
can be approximately described using  
a linear relation $T_{\rm d} \approx k \uptau$, 
where $k = 1.18 \pm 0.07$, 
suggesting that $T_{\rm d}$ 
is reasonably 
characterised by $\uptau$ 
in the explored ranges of the control parameters.
Furthermore, for each dataset obtained using fixed 
$\mu$ and $r_0$,
we note that Fig.~\ref{fig:characteristic}c and Fig.~\ref{fig:characteristic}d
indicate that 
$T_d$ decreases with $U$ for $U<U_{\rm s}$
and plateaus for $U>U_{\rm s}$. 
This implies smaller data deviation from the linear fit in Fig.~\ref{fig:characteristic}e
at higher voltages. 
In other words, the linear 
fit better describes the relation between $T_{\rm d}$
and $\tau$ in the high-voltage regime.

\subsection{Critical conditions for jumping droplets
by EWOD actuation from maximum deformation state}

We now seek for the critical condition 
for droplet detachment by AMS.
We first note that 
the droplet actuation dynamics 
in our study
is kept in the underdamped regime,
as it is the requirement  
to enable detachment \citep{Vo2019}.
This requirement 
also ensures that the excess surface energy 
at the maximum deformation state
is not entirely dissipated by the 
viscous effect during retraction. 
Subsequently,
we follow the energy balance approach 
used to determine 
the jumping conditions for 
droplets actuated by AES \citep{Vo2019} 
to formulate 
the critical condition for droplets to 
detach from the solid substrate
by AMS as
\begin{equation}
\label{eq:condition_1}
\Delta E_{\rm s} \ge E_{\rm v} + E_{\rm cl},
\end{equation}
where $\Delta E_{\rm s} = E_{\rm s}^{\rm m} - E_{\rm s}^{\rm d}$ 
is the difference between the 
surface energy 
at maximum deformation 
$E_{\rm s}^{\rm m}$ 
and the surface energy at 
the detach moment $E_{\rm s}^{\rm d}$; 
$E_{\rm v}$ is
the viscous dissipation 
and $E_{\rm cl}$
contact line elasticity energy 
during retraction. 
We note that 
the kinetic energy of a droplet vanishes
at its maximum deformation state.
The gravitational potential energy is  
negligible compared to the surface energy
as the Bond number ${\rm Bo} = (\rho - \rho_{\rm o}) g r_0^2 \sigma^{-1}$
is small ($2.14 \times 10^{-4} \le \rm{Bo} \le 1.23 \times 10^{-1}$). 
Here, $\rho_{\rm o} = 873 \,{\rm kg\,m^{-3}}$ 
is the 
density of silicone oil and 
$g = 9.81\,{\rm m\,s^{-2}}$ is the gravitational acceleration.

We first focus on the surface energy difference 
$\Delta E_{\rm s}$ 
and note that 
it can be written as 
$\Delta E_{\rm s} = 
E_{\rm s}^{\rm m} - E_{\rm s}^{\rm e} 
+E_{\rm s}^{\rm e} - E_{\rm s}^{\rm d}$, where 
$E_{\rm s}^{\rm e}$ is the surface energy at equilibriated state after 
applying the electrowetting effect without turning it off. 
On one hand, we note that the 
surface energy difference $E_{\rm s}^{\rm e} - E_{\rm s}^{\rm d}$
between 
the equilibrated state and detachment 
was already formulated \citep{Vo2019}:
$E_{\rm s}^{\rm e} - E_{\rm s}^{\rm d} = 
[2(1 + \cos \theta_{\rm e})^{-1}  
- 4 (r_0/r_{\rm e})^2 
- \cos \theta_0] \sigma \pi r_{\rm e}^2$,
where $\theta_{\rm e}$, $r_{\rm e}$ are respectively the contact angle
and contact radius of the droplet at the equilibriated state; 
$\theta_0$ is the contact 
angle of the droplet at the initial state.
Moreover, 
we argue that the surface energy difference 
$E_{\rm s}^{\rm m} - E_{\rm s}^{\rm e}$ 
comes from the capillary wave generated 
at the beginning of droplet actuation;
this wave would not occur if 
the voltage $U$ were to ramp 
up slowly to keep the spreading quasi-static. 
As a result, the 
surface energy difference 
$E_{\rm s}^{\rm m} - E_{\rm s}^{\rm e}$
is calculated by the 
energy 
carried by the capillary wave
$E_{\rm s}^{\rm m} - E_{\rm s}^{\rm e} 
= 0.5 \rho a^2 \omega^2 l S$,
where $a \sim r_0  {\rm We}^{1/2} \sin \theta_0 (T_{\rm p} \omega)^{2/3}/[2\pi (1 - \xi^2)^{1/2}]$
is the wave amplitude 
at time $t = T_{\rm p}$;
$\omega \sim (\sigma/\rho r_0^3)^{1/2}$ 
is the angular frequency of the wave; 
$l \sim r_0$ is the wavelength;
$\xi = \lambda/(\sigma \rho r_0)^{1/2}$
is the decaying ratio 
of the capillary waves; $\lambda$ 
is the contact line friction
coefficient;
$S = 2(1 + \cos \theta_e)^{-1} \pi r_e^2$ 
is the area of a hypothetical 
spherical cap having contact 
angle $\theta_{\rm e}$ 
and base radius $r_{\rm e}$;
and ${\rm We}$ is 
the contact line Weber number,
defined to directly relate 
to the applied voltage $U$ by the relation 
${\rm We} = [(\epsilon \epsilon_0/2 d \sigma)^{1/2} (U - U_{\rm c}) + 1]^2$.
Here, $U_{\rm c}$ 
is the critical voltage 
for capillary 
wave generation on the droplet's surface \citep{Vo2021b}.
In our experiment, 
$U_{\rm c}$ varies from 
$70\,$V to $110\,$V 
depending on the viscosity 
and radius of the actuated droplet.
We therefore obtain the expression of the surface energy difference:
\begin{equation}
\label{eq:Es_1}
\Delta E_{\rm s} = 
\left[\frac{2}{1 + \cos \theta_e}  
- 4 \left(\frac{r_0}{r_e}\right)^2 
- \cos \theta_0  
+ {\rm We} \frac{ (T_{\rm p} \omega)^{4/3}}{4\pi^2(1 - \xi^2)} \frac{\sin^2 \theta_0}{1 + \cos \theta_{\rm e}} \right] \sigma \pi r^2_{\rm e}.
\end{equation}

We note that 
the damping ratio $\xi$
is essentially similar to 
the Ohnesorge number 
Oh $=\mu (\sigma \rho r_0)^{-1/2}$
in which the contact line friction coefficient $\lambda$ 
is used instead of the liquid's viscosity $\mu$ to represent dissipation.
In our analysis, $\lambda$ is obtained 
 empirically using the relation 
 $\lambda = C (\mu \mu_{\rm o})^{1/2}$, 
 where $C = 32.9$ 
 is a fitting parameter \citep{Vo2018,Vo2019}.
By using $\xi$ instead of Oh,
the dissipation 
in the liquid bulk is effectively neglected 
\citep{Vo2019, Carlson2012, Carlson2012a}. 
Indeed, the ratio $\xi/{\rm Oh}$, calculated 
for all of our experiments, 
varies from $5.47$ to $45.3$,
strongly suggesting 
that the dissipation at the contact line is dominant.
As a result, 
we ignore bulk dissipation 
in our estimation of $
E_{\rm s}^{\rm m} - E_{\rm s}^{\rm e}$
to arrive at Eq.~\ref{eq:Es_1}.
Also, by only considering
dissipation at the contact line, 
we obtain an expression for
$E_{\rm v}$ \citep{Vo2019}:
\begin{equation}
\label{eq:Ev}
E_{\rm v} \sim \lambda \frac{r_{\rm m}}{T_{\rm d}} \pi r^2_{\rm m} \approx \lambda \frac{r_{\rm e}}{T_{\rm d}} \pi r^2_{\rm e}.
\end{equation}

The contact line elasticity energy $E_{\rm cl}$
is the surface energy accumulation due 
to pinning and subsequent stretching of 
the liquid-oil interface at the vicinity of 
the contact line \citep{Joanny1984,Vo2019}
and 
is determined by a similar approach with \cite{Vo2019}:
\begin{equation}
\label{eq:Ecl}
E_{\rm cl} \sim \kappa \sigma \pi r^2_{\rm m} \approx \kappa \sigma \pi r^2_{\rm e}.
\end{equation}
Here, $\kappa = \pi \sin^2 \theta_{\rm r}/ {\rm ln} (r_0/\gamma)$,
where $\theta_{\rm r}$ 
is the receding contact angle, 
and $\gamma$ is the defect's size.
In our experiment, 
which was conducted using 
the same setup and materials 
with \cite{Vo2019},
$\theta_{\rm r} \approx 121^{\circ}$ 
and the average defect's size is 
$\gamma \approx 260\,$nm.
The parameter $\kappa$ only changes minutely 
within the explored droplet radius: 
$\kappa = 0.27\pm 0.03$
for $0.08\,{\rm mm} \le r_0 \le 1.25\,{\rm mm}$.

Subtituting Eqs.~\ref{eq:Es_1}, \ref{eq:Ev}, \ref{eq:Ecl}
into Eq.~\ref{eq:condition_1},
the condition at the transition 
between nondetachable and detachable behaviours
is
\begin{equation}
\label{eq:jumpCon1}
\frac{2}{1 + \cos \theta_{\rm e}} - 4\left(\frac{r_{\rm 0}}{r_{\rm e}}\right)^2 - \cos \theta_0 
 + {\rm We} \frac{(T_{\rm p} \omega)^{4/3}}{4\pi^2(1 - \xi^2)} \frac{\sin^2 \theta_0}{1 + \cos \theta_{\rm e}}
= \frac{r_{\rm e} \lambda}{T_{\rm d} \sigma } + \kappa.
\end{equation}

As shown in the previous sections, 
both the activation time $T_{\rm p}$
and the detach time 
$T_{\rm d}$
are well approximated 
by the underdamped characteristic spreading time $\uptau$.
We therefore simplify Eq.~\ref{eq:jumpCon1} 
as 
\begin{equation}
\label{eq:jumpCon2}
\alpha  +  \Psi (\theta_{\rm e}, \xi) {\rm We} = \pi^{-1} \beta \eta^{1/2} \xi + \kappa,
\end{equation}
where $\beta = r_{\rm e}/r_{\rm 0} = (1-\cos \theta_{\rm e}^2)^{1/2} [4 (1-\cos \theta_{\rm e})^{-2} (2 + \cos \theta_{\rm e})^{-1}]^{1/3}$, $\alpha = 2 (1 + \cos \theta_{\rm e})^{-1} - 4 \beta^{-2} - \cos \theta_0$,
and
$\Psi(\theta_{\rm e}, \xi) = 0.25\,\sin^2 \theta_0 (1 + \cos \theta_{\rm e})^{-1} (\pi \eta)^{-2/3} (1 - \xi^2)^{-1}$.
We note that $\theta_{\rm e}$
depends on the electrowetting number $\eta$ 
following the Young-Lippmann's equation 
$\cos \theta_{\rm e} - \cos \theta_{\rm 0} = \eta$ \citep{Mugele2005}.

\begin{figure}
\includegraphics[width=0.5\textwidth]{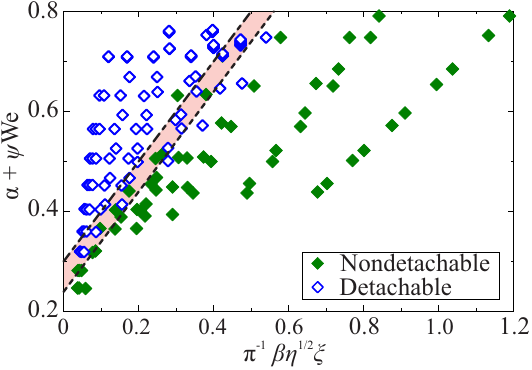}
\centering
\caption{
Plot showing $\alpha + \Psi {\rm We}$ 
versus $\pi^{-1} \beta \eta^{1/2} \xi$
using the data of
the nondetachable 
and detachable behaviours shown
in Fig.~\ref{fig:phaseDiagram}. 
The shaded area indicates 
the transitional extent due to variation in 
$\kappa$ ($\kappa = 0.27\pm 0.03$).
}
\label{fig:PDNormal}
\end{figure}

In Fig.~\ref{fig:PDNormal}, we show 
a plot of $\alpha + \Psi {\rm We}$ versus
$\pi^{-1} \beta \eta^{1/2} \xi$
of all the data obtained for the  
nondetachable
and detachable behaviours shown in 
Fig.~\ref{fig:phaseDiagram}.  
Here, the composite terms
$\alpha + \Psi {\rm We}$ and 
$\pi^{-1} \beta \eta^{1/2} \xi$ respectively 
represent the 
driving energy and the 
energy cost for detachment.
The shaded area 
indicates the 
transitional extent caused by variation in $\kappa$
($\kappa = 0.27\pm 0.03$)
resulted from
variations 
in both the defect's size $\gamma$
and the radius $r_0$.
We note that all the terms in Eq.~\ref{eq:jumpCon2}
are calculated using the system parameters except the contact line friction coefficient, which is determined empirically 
using $\lambda = C (\mu \mu_{\rm o})^{1/2}$
with 
$C = 32.9$ \citep{Vo2018, Vo2019}, 
and the receding contact angle $\theta_{\rm r} \approx 121^{\circ}$ determined independently from our previous work \citep{Vo2019}.
The excellent agreement 
between the 
experimental data and the
formulated jumping condition 
(Eq.~\ref{eq:jumpCon2}) 
thus 
highlights 
our analysis as a predictive tool 
for utilizing modulated electrowetting in droplet actuation and detachment from surfaces. 

\section{Conclusions}
We aim to systematically investigate droplet detachment induced by EWOD using actuation from maximum deformation state (AMS). 
By varying droplet radius, viscosity, and applied voltage,
we demonstrate 
a significant expansion of 
the detachable regime 
for droplets actuated by using the 
electrowetting effect with AMS and provide a comprehensive phase diagram
of the detachment behaviours.
The applied voltage for AMS,
no longer bounded by
contact angle saturation limit,
enables detachment of droplets as small as 80\,$\upmu$m in radius.
This introduces a powerful tool 
for applications requiring 
actuation and detachment of droplets from solid surfaces, 
in particular those 
dealing with small droplets or highly viscous liquids.
We then provide a detailed characterisation of detach velocity and detach time of actuated droplets.
Finally, we developed a 
theoretical prediction for the critical condition 
causing droplet detachment.
The theoretical prediction 
is consistent with our experimental data for 
liquid viscosity ranging from 1\,mPa to 68.7\,mPa, 
droplet size from 0.08\,mm to 1.5\,mm, 
and applied voltage from 60\,V to 200\,V.

We note that our study is 
limited within the droplet-in-oil setting
and further studies may be required
to explore the limit of our analysis in
the droplet-in-air setting, which is typically known 
for stronger
hysteresis effect and electrical 
instability at the three-phase contact line.
Nevertheless, 
our study may serve as a strong basis for 
wider use of electrowetting 
in applications requiring precise actuation of droplets
such as tissue engineering, 
digital microfluidics, and 3D printing.
Our results also provide key insights to 
mechanistic understanding 
of related phenomena such as coalescence-induced 
jumping of droplets
\citep{Boreyko2009,Farokhirad2015,Liu2014e} 
or droplet bouncing on solid substrates \citep{Sanjay2023}.

\section*{Acknowledgments}
This study is supported by Nanyang Technological
University, the Republic of Singapore's Ministry of Education 
(MOE, grant number MOE2018-T2-2-113), 
and
the RIE2020 Industry Alignment Fund -- Industry Collaboration Projects (IAF--ICP) 
Funding Initiative, as well as cash and in-kind contribution from the industry partner, HP Inc.

\section*{Declaration of Interests}
The authors report no conflict of interest.

\bibliographystyle{jfm}


\end{document}